\documentclass[preprint,12pt]{elsarticle}

\usepackage[T1]{fontenc}
\usepackage[utf8]{inputenc}
\usepackage[english]{babel}
\usepackage{courier}
\usepackage{graphicx}
\usepackage{float}
\usepackage{amsthm}
\usepackage{amsmath}
\usepackage{amssymb}
\usepackage{concrete}
\usepackage{geometry}
\usepackage{caption}
\usepackage{subcaption}
\usepackage{algorithm}
\usepackage{algorithmic}
\usepackage{wrapfig}
\usepackage{multirow}
\usepackage{multicol}
\usepackage{epstopdf}
\usepackage{hyperref}
\DeclareFontSeriesDefault[rm]{bf}{bx}

\theoremstyle{plain}

\theoremstyle{definition}
\newtheorem{definition}{Definition}
\theoremstyle{remark}

\floatname{algorithm}{Procedure}

\newcommand{\rank}{\mathrm{rank}\kern 2pt}

\allowdisplaybreaks

\journal{Journal of Computational Physics}

\begin{document}

\title{Low-rank Monte Carlo for Smoluchowski-class equations}

\author[1,2]{A.I.~Osinsky}

\address[1]{Skolkovo Institute of Science and Technology, 
{Bolshoy Boulevard 30, bld. 1, Moscow, Russia 121205}}
\address[2]{Institute of Numerical Mathematics of Russian Academy of Sciences, Moscow, 119333 Russia}


\begin{abstract}
The work discusses a new low-rank Monte Carlo technique to solve Smoluchowski-like kinetic equations. It drastically decreases the computational complexity of modeling of size-polydisperse systems. For the studied systems it can outperform the existing methods by more than ten times; its superiority further grows with increasing system size. Application to the recently developed temperature-dependent Smoluchowski equations is also demonstrated.
\end{abstract}

\maketitle


\section{Introduction}\label{int-sec}

The present study addresses the fast methods of solving the system of Smoluchowski coagulation equations
\begin{equation}\label{n-eq}
\frac{d}{{dt}}{n_k} = \frac{1}{2}\sum\limits_{i + j = k} {{C_{ij}}{n_i}{n_j}}  - \sum\limits_{j = 1}^\infty  {{C_{kj}}{n_k}{n_j}}, \quad k = \overline {1,\infty } 
\end{equation}
and it's analogues. Here the sizes of the aggregating objects  are assumed to be discrete, ranging in size from 1 to infinity. $n_k$ denote the number densities of particles (aggregates) of size $k$, that is, of the aggregates comprising $k$ elementary units -- the monomers. The system \eqref{n-eq} describes the rate of change of number densities $n_k$ depending on how often particles of size $i$ and $j$ aggregate (every collision is assumed to lead to aggregation). The collision rate is determined by values of a kernel elements $C_{ij} = C_{ji}$.

Recently, new methods for fast solution of Smoluchowski ODE were developed \cite{matveev,matveev_CPC2018,gen-smol}. They are based on the low-rank approximation to the collision kernel $C$. Since in practice it is impossible to numerically solve the infinite number of equations \eqref{n-eq}, their number is explicitly bounded by $M$:
\begin{equation}\label{eq:nfinite}
\frac{d}{{dt}}{n_k} = \frac{1}{2}\sum\limits_{i + j = k} {{C_{ij}}{n_i}{n_j}}  - \sum\limits_{j = 1}^M {{C_{kj}}{n_k}{n_j}}, \quad k = \overline {1,M} 
\end{equation}
Thus, $M$ is the largest aggregate size (mass) in the system. Solving these $M$ equations directly requires $O \left( M^2 \right)$ operations every time step. Low-rank approximation allows to perform each time step in only $O \left(MR \log M \right)$ operations, where $R$ is the approximation rank, which is usually much lower than $M$. Thus, it is very efficient, when $R \ll M / \log M$.

However, not all systems of aggregation equations can be solved this way. Although all popular kernels (ballistic \cite{BallOrig,BallDimen,BallMolec}, Brownian \cite{BrownScal,BrownGeneral,BrownGeneral2}, etc.) are low-rank or can be approximated by low-rank kernels, problems arise when gelation can happen \cite{van-dongen}, after which point the ODE solution breaks down.
Moreover, the conservation of mass may require very large number of equations $M$ to exploit a finite system, which approximates the infinite system well. Finally, in the course of time, one usually needs to appropriately change the time step, as the evolution of the system greatly affects the rate of change of the number densities.

While the aforementioned problems can be solved using special tricks, there is another simple approach that can bypass them. Namely, instead of solving the ODE equations, one can perform Monte Carlo simulations with a large, but finite number of particles, with $C_{ij}$ defining the collision rates for the particles of the corresponding sizes. In this case, the mass conservation is satisfied automatically, one can easily study gelation, and the time step can be automatically defined by the time between collision events. Unfortunately, the conventional Monte Carlo methods require a lot of computation time to choose the colliding particles.

Modern Monte Carlo methods \cite{monteweighted,montemanyevents,montegel,montefrac} are based on the classical inverse \cite{gillespie,inverse} or accepta\-nce-rejection \cite{monte-random} methods. Modern research mostly answers the questions, what method should be used for a particular system \cite{montesurvey}, how to parallelize the methods \cite{montefrac,comparison,inversepar} or perform the generalization the classical methods to apply them to more and more complex processes \cite{montemanyevents,MonteCarlo2003}. However, the underlying principle of selecting the colliding pair of particles stays the same, and it is usually the bottleneck in terms of computational complexity.

Here we present the modification of the Monte Carlo method, which exploits the low-rank property to reduce the required number of operations for each collision. Moreover, the low-rank approach does not require the kernel itself to be low-rank or have a good low-rank approximation; instead, it is only needed to be bounded from above by some low-rank kernel $\tilde C$, $\rank \tilde C = R$. Note that any matrix $C$ can be bounded by a low-rank matrix $\tilde C$. For instance, consider the kernel $\tilde C$ with the elements:
\begin{equation}
  \tilde C_{ij} = \max\limits_{1 \leqslant k,l \leqslant M} C_{kl}.
\end{equation}
Then $\tilde C_{ij}$ is a constant and thus a rank 1 kernel. This bound is used by the classical acceptance-rejection method, which replaces collision rates by $\tilde C_{ij}$ and introduces the rejection step to compensate for the overestimate. Low-rank method generalizes this approach, allowing to use any low-rank kernel $\tilde C$. The points is that when $\tilde C$ is close to the original kernel $C$, the rejection probability is reduced. If $\tilde C$ differs from $C$ at most by a constant factor, the cost of each collision step in the low-rank method is $O(R \log M)$. On the other hand, the classical methods are polynomial in the maximum size $M$. Although maximum particle size $M$ usually grows logarithmically with an increasing number of particles $N$ (due to the exponential tail of the size distribution), it grows very quickly with increasing system time during aggregation. Thus, studying aggregation processes at large time intervals requires much more efficient algorithms than currently available.

Whether there exists a low-rank kernel $\tilde C$ close to $C$ depends on the exact expression for the kernel $C$. For instance, any kernel $C_{ij} = f(i)f(j)$ is, by definition, a rank 1 kernel, but, unfortunately, there is no universal way to check whether the kernel can be bound by a low-rank one. Therefore, we focus on the most widely used collision mechanisms (Brownian and ballistic) and show that all collision kernels corresponding to them can be efficiently bound by low-rank kernels with ranks $R = 2$ and $R = 3$ respectively.

The proposed approach 
can be also used in temperature-dependent models \cite{nature,bodrova}, where the authors introduce partial temperatures $T_i$, which need to be updated after each collision. It was previously successfully used to construct the phase diagram for temperature-dependent Smoluchowski equations \cite{diagram} and check numerically the exact analytical solutions \cite{meexact}. Here we explain this method in general.

The paper is organized as follows.

In the next section \ref{sec-conv} the conventional Monte Carlo methods for Smoluchowski equations are described.

In section \ref{sec-smol} we demonstrate how to apply the new method to conventional Smoluchowski equations.

In section \ref{sec-temp} temperature-dependent Smoluchowski equations are studied, and it is shown that they can also be easily solved with the new method.

The conclusions are presented in the last section \ref{sec-con}.

\section{Conventional methods}\label{sec-conv}

Equations \eqref{n-eq} and \eqref{eq:nfinite} are mean-field equations, dealing with continuum quantities $n_k$ -- number densities of clusters of size $k$, defined as $n_k = N_k/V$, where $V$ is the system volume and $N_k$ -- the number of clusters of size $k$ in this volume. These equations ignore the discrete nature of particles and thus unavoidable fluctuations. In contrast, Monte Carlo approach allows to model the kinetic processes with fluctuations and thus reflects more adequately real aggregation processes. 

Let $N = \mathop \sum\limits_{i = 1}^M N_i$ be the total number of particles, where $M$ is the maximum particle mass in the system. $C_{ij}$ denote the collision rate of particles of size $i$ and $j$. Then, if a collision happens, one can say that the probability that it was between the particles of size $i$ and $j$ is \cite{inverse,analytic}
\begin{equation}
  P_{ij} = \frac{C_{ij} N_i N_j}{\mathop \sum\limits_{i, j} C_{ij} N_i N_j}.
\end{equation}
Monte Carlo method is performed as follows:
\begin{enumerate}
\item Initialize the system with $N_i$ particles of size $i$. Set $t = 0$.
\item Find the time interval $\Delta t$ until the next aggregation event. Set $t := t + \Delta t$.
\item Select the sizes, according to probabilities $P_{ij}$.
\item Perform aggregation:
\begin{equation}\label{eq:aggstep}
N_i := N_i - 1, \quad N_j := N_j - 1, \quad N_{i+j} := N_{i+j} + 1.
\end{equation}
\item Go to step 2.
\end{enumerate}

In general, step 4 can also account for aggregation probability if not all collisions lead to aggregation and some other types of interaction like fragmentation \cite{smolfrag}, shattering \cite{mematveev} or update of partial temperatures both during aggregation and bouncing collisions \cite{diagram}. Here we will focus on the aggregation process.

A naive way of choosing probabilities goes through all $M^2$ possible values of $i$ and $j$, which is very inefficient. Several conventional methods use different approaches to reduce this number and determine the time interval $\Delta t$. We briefly describe them below.

\textbf{Gillespie method} was initially developed for chemical reactions \cite{gillespie}. In the case of aggregation, it's various modifications are called ``inverse methods'' \cite{inverse}. We are going to compare our algorithm with the version from \cite{inversegood}, which uses an array of sizes instead of a (larger) array of particles and thus was proven to be more efficient (see, e.g. \cite{montesurvey}); it can be described as follows. Denote
\begin{equation}
\begin{aligned}
  s_i & = \sum\limits_{j=1}^M C_{ij} N_i N_j, \\
  S & = \sum\limits_{i=1}^M s_i.
\end{aligned}
\end{equation}
The size $i$ of the first aggregation partner is selected with probability $s_i/S$ and after that size $j$ particle is selected with probability $C_{ij} N_i N_j / s_i$. Therefore, one first determines the size $i$ of the first particle and then the size $j$ of the second one. Thus, at most $2M$ checks to determine the colliding particles and so is not fast for large $M$. If, instead, one uses an array of particles, then up to $2N$ checks are required. The main cost, however, comes from the updates of the rates $C_{ij}$ as three rows and columns should be updated after each collision (corresponding to sizes $i$, $j$ and $i+j$). More details can be found in \cite{inversegood}.

There exist other methods, which update time differently using a fixed time step \cite{montetime}, but it increases the complexity compared to other techniques \cite{montesurvey}. One can also sort the values of $s_i$ to choose the first particle faster. Other versions of the inverse method use the array of particles instead of the array of sizes, which leads to the total quadratic $O \left( N^2 \right)$ complexity of the algorithm \cite{comparison}, which is not desirable.

The time step follows from the assumption that the collisions are a Poisson process (where probability of the collision event is proportional to the duration on a small time interval \cite{analytic}) and the time step is generated from the exponential distribution as follows \cite{inverse}:
\begin{equation}\label{eq:deltat}
  \Delta t = - \Delta t_{\rm avg} \ln \left( {\rm rand} (0, 1] \right).
\end{equation}
The average expected time $\Delta t_{\rm avg}$ is inversely proportional to the total rate of collisions, which follows from summing up all the equations in \eqref{eq:nfinite}:
\begin{equation}\label{eq:deltatavg}
  \Delta t_{\rm avg} = \frac{2V}{\sum\limits_{i,j} C_{ij} N_i N_j} = 2V/S,
\end{equation}
where $V = N / \sum\limits_{i=1}^M n_i$ is the system volume. The derivation of the average expected time step for both inverse and acceptance-rejection methods can be found in \cite{mematveev}.

\textbf{Acceptance-rejection method.} Another technique is based on the random choice of the colliding particles \cite{monte-random}. The collision is then accepted with the probability
\begin{equation}\label{eq:accrej}
  p_{ij} = \frac{C_{ij}}{C_{\max}},
\end{equation}
where $C_{\max} = \max\limits_{1 \leqslant k,l \leqslant M} C_{kl}$, and rejected otherwise. The number of checks until eventual acceptance depends on how close the average (over all pairs of particles) value of $C_{ij}$ is to the maximum $\max\limits_{1 \leqslant k,l \leqslant M} C_{kl}$. As $C_{ij}$ usually contain some small positive powers of $i$ and $j$, the number of checks is $O \left( M^{\alpha} \right)$ with $\alpha$ depending on the scaling properties \cite{scaling} of a particular collision kernel. In the low-rank method we will use a similar idea, replacing $C_{\max}$ by an arbitrary low-rank bound $\tilde C_{ij}$.

In acceptance-rejection method we can't use the same equations \eqref{eq:deltat}-\eqref{eq:deltatavg} as the total rate $S$ is unknown. Therefore, the time is instead updated every time the sizes are randomly chosen (independent on whether there were accepted or not) and equation \eqref{eq:deltatavg} is replaced by
\begin{equation}
  \Delta t_{\rm avg} = \frac{2V}{C_{\max} N (N-1)},
\end{equation}
which leads to the same time step distribution after acceptance \cite{mematveev}.

From the algorithmic perspective, the described inverse and acceptance-rejection methods are completely equivalent since they compute the same collision probabilities and the time step distributions are also equal. The only difference is in the computation time. The aim of the present study is reduce the respective complexity by providing a more efficient way to choose the particle pairs. There are also ways to change the time step or introduce weighted particles, which changes not only the cost, but also the accuracy of the simulation \cite{monteweighted,multimonte}.

\section{New Monte Carlo method}\label{sec-smol}

In this section an application of low-rank approximations to Monte Carlo is discussed.

We start with a reminder of the notion of the matrix rank.

\begin{definition}
Let $A \in \mathbb{R}^{m \times n}$ be an $m \times n$ matrix with real entries $A_{ij}$. The minimum value of $R$ for which $A$ can be expressed as a sum of outer (dyadic) products
\begin{equation}\label{eq-rank}
  A = \mathop \sum\limits_{k = 1}^R \vec u^k \otimes \vec v^k
\end{equation}
of two sets of vectors
\begin{equation}
\begin{aligned}
  \vec u^k \in \mathbb{R}^m, & \quad k = \overline{1, R}, \\
  \vec v^k \in \mathbb{R}^n, & \quad k = \overline{1, R}
\end{aligned}
\end{equation}
is called the \textbf{rank} of matrix $A$.
\end{definition}

When $R \ll \min \left( m, n \right)$, expression (\ref{eq-rank}) is called \textbf{low-rank decomposition} of matrix $A$. For example, a product kernel $C_{ij} = ij$ has rank $1$ even after multiplication by number densities $N_i$, since it can be expressed as
\begin{equation}
  C_{ij} N_i N_j = u_i^1 \cdot v_j^1
\end{equation}
with
\begin{equation}
  u_i^1 = i N_i, \quad v_j^1 = j N_j.
\end{equation}

The reason to use low-rank decomposition is that it provides a simpler way to choose the sizes of colliding particles in Monte Carlo. Let us first consider the case, where matrix $C$ itself is low-rank (we will return to the case $C_{ij} \leqslant \tilde C_{ij}$ with $\tilde C_{ij}$ being low-rank later). For example, let $R = 1$ so that $C_{ij} N_i N_j = u_i \cdot v_j$. Then the probability that the first particle has size $i$ can be calculated as
\begin{equation}\label{eq-pi}
  p_i = \frac{\mathop \sum\limits_{j = 1}^M C_{ij} N_i N_j}{\mathop \sum\limits_{i, j = 1}^M C_{ij} N_i N_j} = \frac{\mathop \sum\limits_{j = 1}^M u_i v_j}{\mathop \sum\limits_{i, j = 1}^M u_i v_j} = \frac{u_i \mathop \sum\limits_{j = 1}^M v_j}{\mathop \sum\limits_{i = 1}^M u_i \mathop \sum\limits_{j = 1}^M v_j} = \frac{u_i}{\mathop \sum\limits_{i = 1}^M u_i}
\end{equation}
and depends only on the components of vector $\vec u$. Similarly,
\begin{equation}\label{eq-pj}
  p_j = \frac{v_j}{\mathop \sum\limits_{j = 1}^M v_j}
\end{equation}
depends only on the components of vector $\vec v$.

If the rank is greater than 1, the matrix $C_{ij} N_i N_j$ can be represented as a sum of rank 1 kernels (\ref{eq-rank}), which can be considered separately. To determine which rank 1 kernel is to be used, the total collision rates are compared, and $k$-th outer product $\vec u^k \otimes \vec v^k$ is chosen with the probability
\begin{equation}
  p_k = \frac{\mathop \sum\limits_{i, j = 1}^M u_i^k v_j^k}{\mathop \sum\limits_{k = 1}^R \mathop \sum\limits_{i, j = 1}^M u_i^k v_j^k} = \frac{\mathop \sum\limits_{i = 1}^M u_i^k \mathop \sum\limits_{j = 1}^M v_j^k}{\mathop \sum\limits_{k = 1}^R \left( \mathop \sum\limits_{i = 1}^M u_i^k \mathop \sum\limits_{j = 1}^M v_j^k \right) }.
\end{equation}

So far, the advantages of the low-rank approach are not yet apparent. Now we demonstrate it's advantages.

\begin{wrapfigure}{R}{0.5\textwidth}
\centering
\includegraphics[width=0.5\columnwidth]{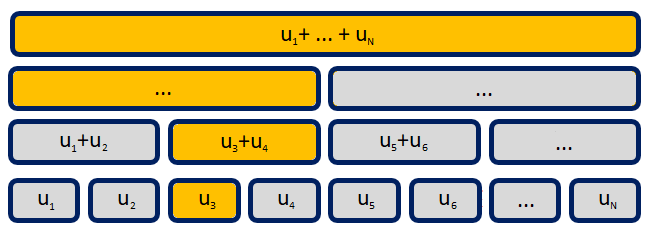}
\caption{Segment tree for vector $\vec u$.
}
\label{fig-tree}
\end{wrapfigure}

The task is to speed up the choice of particle sizes according to the probabilities $p_i$ and $p_j$ (\ref{eq-pi}, \ref{eq-pj}), which form arrays of size $M$. It can be done using the segment tree data structure (see figure \ref{fig-tree}).

The segment tree is a binary tree, which contains partial sums of the array elements. Whenever a random number
\begin{equation}
  r := {\rm rand}(0, 1] \cdot \mathop \sum\limits_{i = 1}^M u_i
\end{equation}
is chosen to determine the size of the first particle, the size can be chosen in $O \left( \log M \right)$ steps by subtracting partial sums of $u_i$ from $r$. Moreover, segment tree can be updated in $O \left( \log M \right)$ operations by updating the partial sums from bottom (leaf) to the top (root). As there are $R$ segment trees to be updated, the total complexity is $O \left( R \log M \right)$. If $R$ does not rise with increasing maximum particle size $M$, the asymptotic complexity is lower than $O \left( M^{\alpha} \right)$ for any $\alpha > 0$, which makes it superior to both inverse and acceptance-rejection methods.

In the case, when the kernel $C$ is not low-rank, the acceptance-rejection strategy similar to \eqref{eq:accrej} can be applied. Namely, the kernel $C$ can be bounded from above by another kernel $\tilde C$, which is low-rank. For example, the ballistic kernel \cite{BallOrig,BallDimen,BallMolec}
\begin{equation}
  C_{ij} = {\rm const} \cdot \left( r_i + r_j \right)^2 \sqrt{\frac{T_i}{m_i} + \frac{T_j}{m_j}},
\end{equation}
where $r_i$ are cluster radii, $T_i$ are partial temperatures, $m_i$ are masses. It can be approximated from above by the following one:
\begin{equation}
  \tilde C_{ij} = {\rm const} \cdot \left( r_i + r_j \right)^2 \left( \sqrt{\frac{T_i}{m_i}} + \sqrt{\frac{T_j}{m_j}} \right).
\end{equation}

The above kernel $\tilde C$ as well as the combination $\tilde C_{ij} N_i N_j$ have rank $6$, which can be seen by opening the brackets. To account for the fact that higher collision rates are used, an additional acceptance-rejection step is added. Namely, the collision of particles $i$ and $j$ is accepted with the probability
\begin{equation}
  p_{ij} = \frac{C_{ij}}{\tilde C_{ij}} = \frac{\sqrt{\frac{T_i}{m_i}} + \sqrt{\frac{T_j}{m_j}}}{\sqrt{\frac{T_i}{m_i} + \frac{T_j}{m_j}}}.
\end{equation}

As the collision kernel $\tilde C$ is symmetric, one can also use a non-symmetric upper bound matrix $A$ such that
\begin{equation}\label{eq:tildeca}
  \tilde C_{ij} = A_{ij} + A_{ji}.
\end{equation}
In the case of the ballistic kernel, one can choose
\begin{equation}
  A_{ij} = \left( r_i + r_j \right)^2 \sqrt{\frac{T_i}{m_i}},
\end{equation}
which significantly decreases the rank as $\rank A = 3$, while $\rank \tilde C = 6$. This leads to different probabilities of selecting the ordered pairs $i > j$ and $j > i$, but since the collision does not depend on the order of particles in the pair and the total (unordered) rate is unchanged, there is no error.

In general, one can apply the same procedure: opening the brackets, bounding the powers/roots and using symmetry to construct efficient upper bounds for a wide range of kernels, given their analytical expressions.

Classical dimensionless ballistic kernel \cite{monte-random} 
\begin{equation}\label{eq-balcon}
  C_{ij} = \left( i^{1/3} + j^{1/3} \right)^2 \sqrt{\frac{1}{i} + \frac{1}{j}},
\end{equation}
which will be used in simulations, similarly has an upper bound
\begin{equation}
  A_{ij} = \left( i^{1/3} + j^{1/3} \right)^2 \sqrt{\frac{1}{i}}.
\end{equation}

Overall, the discussed algorithm reads:

\paragraph*{Low-rank Monte Carlo algorithm}
\begin{algorithmic}[1]
\REQUIRE{
Initial maximum particle size $M$.

Vector of initial numbers of particles $N_i, \quad i = \overline{1, M}$.

Initial volume $V = N / \sum\limits_{i = 1}^M n_i$.

Matrix $A \in \mathbb{R}^{M \times M}$ in the form $A_{ij} N_i N_j = \sum\limits_{k = 1}^R \vec u_i^k \otimes \vec v_j^k, \quad \vec u^k, \vec v^k \in {\mathbb{R}^M}$, such that $C_{ij} \leqslant \tilde C_{ij} = A_{ij} + A_{ji}$. 

Final time $maxtime$.

}
\ENSURE{$n_i = N_i/V$ is equal to the number densities of the size-$i$ particles at time $t = maxtime$.
}

\STATE \COMMENT{Set initial number of particles $N_{0}$:}
\STATE $N_{0} := \sum\limits_{i = 1}^M {N_i}$
\STATE \COMMENT{Set current number of particles $N$ equal to initial number of partiles:}
\STATE $N := N_{0}$
\STATE Create segment trees $u_t^k$ and $v_t^k$ based on vectors $\vec u^k$ and $\vec v^k$
\STATE $curtime := 0$
\WHILE{$curtime < maxtime$}
  \REPEAT
    \STATE \COMMENT{First elements of segment trees contain the total sum, which can be used to compute the total aggregation rate:}
    \STATE $total\_rate := \sum\limits_{k = 1}^R (u_t^k)_1 (v_t^k)_1$
    \STATE \COMMENT{Update current time:}
    \STATE $curtime := curtime - 2 V \ln \left( {\rm rand} (0,1] \right) / total\_rate$
    \STATE \COMMENT{Choose a rank 1 component:} \label{r1c-line}
    \STATE $r := total\_rate \cdot {\rm rand}(0,1]$
    \FOR{$k := 1$ \TO $R$}
      \STATE \COMMENT{First elements of segment trees contain the total sum, which is subtracted from $r$:}
      \STATE $r := r - (u_t^k)_1 (v_t^k)_1$
      \IF {$r \leqslant 0$}
        \STATE \textbf{break}
      \ENDIF
    \ENDFOR
    \STATE Choose sizes $i$ and $j$ through binary search in segment trees $u_t^r$ and $v_t^r$
    \STATE \COMMENT{Rejection sampling:}
    \IF{$\tilde C_{ij} \cdot {\rm rand}(0, 1] > C_{ij}$}
      \STATE \textbf{continue}
    \ENDIF
    \STATE \COMMENT{Reject collision with itself:}\label{line:itself}
    \IF{($i = j$) \AND ($N_i \cdot {\rm rand}(0,1] \leqslant 1$)} 
      \STATE \textbf{continue}
    \ENDIF
  \UNTIL{$i$ and $j$ are chosen}
  \STATE \COMMENT{Process aggregation:}
  \IF{$i + j > M$} \label{line-ninc}
    \STATE $M := 2M$
    \STATE Increase the sizes of all the arrays correspondingly and initialize new values by $0$
  \ENDIF
  \STATE $N_i := N_i - 1$
  \STATE $N_j := N_j - 1$
  \STATE $N_{i+j} := N_{i+j} + 1$
  \STATE $N := N - 1$
  \STATE \COMMENT{Number of particles doubling:}\label{line:doub}
  \IF{$N \leqslant N_{0}/2$}
    \STATE $N := 2N$
    \STATE $V := 2V$
  \ENDIF
  \STATE Update segment trees
\ENDWHILE
\end{algorithmic}

If the sizes $i$ and $j$ are equal (line \ref{line:itself}), there is a $1$ in $N_i$ chance the same particle is selected. Since a single particle can't collide with itself, such case should be rejected. 

In the line \ref{line:doub} a doubling strategy \cite{montetime} is used to keep the number of particles approximately the same, although there exist other ways to keep the total number of particles the same after every aggregation event \cite{montenumber}.

In the described algorithm, the positiveness of all rank 1 components is required, which is usually the case. However, this is not a necessary assumption. If the approximation is not entirely positive, one can skip the choice of the component $k$ (line \ref{r1c-line}) and search in all of the components simultaneously, choosing size $i$ with probability, proportional to
\begin{equation}
  p_i \sim \sum\limits_{k = 1}^R {u_i^k v_{\Sigma}^k}, \quad v_{\Sigma}^k  = \sum\limits_{j = 1}^M {v_j^k} \in {\mathbb{R}^R}.
\end{equation}
The total complexity does not change since the complexity of the choice becomes $O(R \log M)$.

In short, the acceleration of the low-rank Monte Carlo technique comes from the fact that there are rarely any rejections (or there are none if $C = \tilde C$, except when the same particle is chosen), compared to the acceptance-rejection method. And that the binary search is used to select the sizes, instead of the full search, like in the inverse method.

\subsection{Numerical examples}

The new method will be tested on a linear kernel $C_{ij} = i + j$, ballistic kernel (\ref{eq-balcon}) and Brownian kernel \cite{BrownScal}
\begin{equation}
  C_{ij} = \left( i^{1/3} + j^{1/3} \right) \left( i^{-1/3} + j^{-1/3} \right).
\end{equation}
$C$ here is already a low-rank kernel (it has rank $3$):
\begin{equation}
  C_{ij} = i^{1/3} j^{-1/3} + 2 + i^{-1/3} j^{1/3},
\end{equation}
so we can use $\tilde C = C$.

To satisfy \eqref{eq:tildeca}, one can choose
\begin{equation}
  A_{ij} = 1 + \left( i / j \right)^{1/3}.
\end{equation}
Clearly, $\rank A = 2$.

In general, any Brownian kernel of the form
\begin{equation}
  C_{ij} = {\rm const} \cdot \left( r_i + r_j \right) \left( D_i + D_j \right)
\end{equation}
has rank at most $4$, which can be easily seen by opening the brackets. Here $r_i$ and $r_j$ are cluster radii (for fractal clusters they can be arbitrary functions of masses $i$ and $j$) and $D_i$ and $D_j$ are diffusion coefficients for size $i$ and $j$ clusters. We can also use a non-symmetric upper bound (which in this case is exact):
\begin{equation}
  C_{ij} = A_{ij} + A_{ji}, \quad A_{ij} = {\rm const} \cdot r_i \left( D_i + D_j \right)
\end{equation}
with $\rank A = 2$. 

Now we move to the simulation results. Let us start with monodisperse initial condition $n_{i} \left( 0 \right) = \frac{N_i \left( 0 \right)}{V} = \delta_{i1}$. The low-rank technique is compared with the inverse and acceptance-rejection methods. Since eventually all three methods lead to the same probabilities, proportional to $C_{ij}$, and everything else is also the same, the three methods have the same accuracy and even the same distribution of the stochastic error. The only difference is the computation time, which is presented in the tables below. 

The results are presented in table \ref{tab-brown} for Brownian kernel, table \ref{tab-ball} for ballistic kernel and table \ref{tab-lin} for the linear kernel. Note that value of $M$ is doubled every time, when a particle with size larger than $M$ appears. Thus, the size $M$ of the array of particle numbers $N_i$ is always a power of $2$.

\begin{table}[ht]
\caption{Monte Carlo computation time for $10^7$ particles, Brownian kernel.}
\label{tab-brown}
\vspace{-0.3cm}
\begin{center}
\small
\begin{tabular}{|c|c|c|c|c|c|}
\hline
Time, 
$t$ & $M$ & Inverse method, sec & Acceptance-rejection, sec & Low-rank, sec & Collisions \\
\hline
10 & 512 & 67 & 8.3 & 3.5 & $2.3 \cdot 10^7$ \\
\hline
100 & 4096 & 3366 & 18.5 & 7.1 & $3.9 \cdot 10^7$ \\
\hline
1000 & 65536 & -- & 37 & 11 & $5.5 \cdot 10^7$ \\
\hline
10000 & 524288 & -- & 74 & 17 & $7.2 \cdot 10^7$ \\
\hline
100000 & 4194304 & -- & 147 & 27 & $8.9 \cdot 10^7$ \\
\hline
\end{tabular}
\end{center}
\end{table}

Since the Brownian kernel is very close to a constant, the acceptance-rejection method rarely makes rejections and so is not too much slower than low-rank. Still, one can observe the computation time growing as $\sim M^{1/3}$, while for the low-rank method, computation time is proportional to $\log M$ (times the total number of collisions). For the ballistic kernel the computation time for acceptance-rejection method grows $\sim M^{1/2}$ (the exponent $\alpha$ coincides with the absolute value of the scaling parameter $\left| \mu \right|$ \cite{scaling}). Moreover, the acceptance-rejection significantly underperforms on the linear kernel, where the computation time grows linearly with maximum size $M$. Thus, the low-rank technique is the most advantageous for fast growing kernels or when the maximum size $M$ is large.

The inverse method loses to both since it always has linear computation time. The only exception happens for the linear kernel, where acceptance-rejection method also has the complexity $O \left( M \right)$ per collision. Symbol ``--'' is put in the tables when the time limit of 24 hours is exceeded.

\begin{table}[ht]
\caption{Monte Carlo computation time for $10^7$ particles, ballistic kernel.}
\label{tab-ball}
\vspace{-0.3cm}
\begin{center}
\small
\begin{tabular}{|c|c|c|c|c|c|}
\hline
Time, 
$t$ & $M$ & Inverse method, sec & Acceptance-rejection, sec & Low-rank, sec & Collisions \\
\hline
1 & 128 & 6.5 & 5.4 & 4.4 & $1.1 \cdot 10^7$ \\
\hline
5 & 512 & 84 & 17 & 11 & $2.3 \cdot 10^7$ \\
\hline
50 & 8192 & 7967 & 62 & 23 & $4.3 \cdot 10^7$ \\
\hline
500 & 131072 & -- & 221 & 38 & $6.3 \cdot 10^7$ \\
\hline
5000 & 2097152 & -- & 840 & 58 & $8.3 \cdot 10^7$ \\
\hline
\end{tabular}
\end{center}
\end{table}

\begin{table}[ht]
\caption{Monte Carlo computation time for $10^7$ particles, linear kernel.}
\label{tab-lin}
\vspace{-0.3cm}
\begin{center}
\small
\begin{tabular}{|c|c|c|c|c|c|}
\hline
Time, 
$t$ & $M$ & Inverse method, sec & Acceptance-rejection, sec & Low-rank, sec & Collisions \\
\hline
1 & 128 & 8.4 & 14 & 0.89 & $7.6 \cdot 10^6$ \\
\hline
3 & 8192 & 4887 & 172 & 3.5 & $2.2 \cdot 10^7$ \\
\hline
6 & 2097152 & -- & 2370 & 6.7 & $4.4 \cdot 10^7$ \\
\hline
\end{tabular}
\end{center}
\end{table}

\section{Temperature-dependent aggregation}\label{sec-temp}

The described approach can also be used in the case of non-equilibrium temperature distribution.

Let particles of size $i$ have temperature $T_i$. In the temperature-dependent model \cite{PNAS,nature}, the Smoluchowski equations are supplemented by the equations for temperatures, which read:
\begin{equation}\label{nt-eq}
\frac{d}{{dt}}{n_k}{T_k} = \frac{1}{2}\sum\limits_{i + j = k} {{B_{ij} \left( T_i, T_j \right)}{n_i}{n_j}}  - \sum\limits_{j = 1}^\infty  \left( {D_{kj}^{\rm agg} \left( T_i, T_j \right) + D_{kj}^{\rm res} \left( T_i, T_j \right)} \right) {n_k}{n_j} ,\quad k = \overline {1,\infty } ,
\end{equation}
where the kernels $B$, $D$ and the collision kernel $C$ now also depend on temperatures.

In general, not all collision lead to aggregation, so the collision kernel in \eqref{n-eq} should be replaced by aggregation kernel $C_{ij}^{\rm agg}$.

The kernel $B$ accounts for the temperature change of size $k$ particles, which appear from the aggregation of size $i$ and $j$ particles.

The kernel $D^{\rm agg}$ accounts for the temperature change of size $k$ particles when they disappear by aggregation with size $j$ particles.

The kernel $D^{\rm res}$ accounts for the temperature change of size $k$ particles when they collide with particles of size $j$ but do not aggregate. This case is called restitution.

Note that the system \eqref{nt-eq} can be used to study the temperature distribution in polydisperse non-aggregating granular gases \cite{twogas,polygas,bodrova2014}, in which case $B = D^{\rm agg} = 0$ and $C^{\rm agg} = 0$.

In the case of ballistic aggregation with unit masses and particle diameters, the aggregation kernel $C^{\rm agg}$ is defined as
\cite{BallAggFrag, nature}
\begin{equation}\label{eq-c}
  C_{ij}^{\rm agg} = C_{ij} P \left( i, j, T_i, T_j \right) = \sqrt{\pi / 2} \left( i^{1/3} + j^{1/3} \right)^2 \sqrt{\frac{T_i}{i} + \frac{T_j}{j}} P \left( i, j, T_i, T_j \right),
\end{equation}
where $C$ is the collision kernel and the function $P \left( i, j, T_i, T_j \right)$ defines the aggregation probability. The cited models assume the distribution of speeds to be approximately Maxwellian for each particle size before and after each collision. 

In case of temperature-dependent Smoluchowski equations the aggregation step of the Monte Carlo method \eqref{eq:aggstep} also accounts for the changes in partial temperatures. When aggregation happens, the following equations are used to update the temperatures \cite{diagram,meexact}:
\begin{equation}\label{aggtemp-eq}
\begin{gathered}
  \left\{ \begin{gathered}
  T_i := T_i + \frac{T_i - D_{ij}^{\rm agg}/C_{ij}^{\rm agg}}{{{N_i} - 1}} \hfill \\
  T_j := T_j + \frac{T_j - D_{ij}^{\rm agg}/C_{ij}^{\rm agg}}{{{N_j} - 1}} \hfill \\
  T_{i+j} := T_{i+j} + \frac{{B_{ij}/C_{ij}^{\rm agg} - T_k}}{{{N_{i+j}} + 1}} \hfill \\ 
\end{gathered}  \right. \hfill
\end{gathered} 
\end{equation}
And when there is no aggregation, particle numbers are not updated, and the temperatures are updated as follows:
\begin{equation}\label{restemp-eq}
\begin{gathered}
  \left\{ \begin{gathered}
  T_i = T_i - \frac{{D_{ij}^{\rm res}}}{N_i C_{ij} (1 - P)} \hfill \\
  T_j = T_j - \frac{{D_{ij}^{\rm res}}}{N_j C_{ij} (1 - P)} \hfill \\ 
\end{gathered}  \right. \hfill
\end{gathered} 
\end{equation}

These updates are derived the same way temperature-dependent Smoluchowski equations are derived \cite{nature}. In short, the values $C_{ij}$, $C_{ij}^{\rm agg}$, $B_{ij}$, $D_{ij}^{\rm agg}$ and $D_{ij}^{\rm res}$ are calculated by integrating over all possible speeds of size $i$ and $j$ particles, using the corresponding Maxwell distributions. Thus, they describe the number of collisions, aggregation events and changes in temperature in a unit time. If one then divides the change in partial temperatures in a unit time by the number of collisions in a unit time, then the temperature change per collision is obtained, see \cite{diagram}.

Like in the classical case, ballistic kernel $C$ with the elements
\begin{equation}\label{eq-hatc}
C_{ij} = \sqrt {\pi / 2} {\left( {{i^{1/3}} + {j^{1/3}}} \right)^2} \sqrt{\frac{T_i}{i} + \frac{T_j}{j}}
\end{equation}
is the collision rate, and describes the number of collisions in a unit of time. As before, this kernel is used to determine the sizes of the colliding particles. After the decision about sizes, one proceeds with the aggregation event with the probability $P \left( i, j, T_i, T_j \right)$ and with the restitution with the probability $1 - P \left( i, j, T_i, T_j \right)$. The value of $P_{ij}$ as well as the equations for all temperature-dependent kernels are given in \ref{mod:app}.


Similarly to the constant temperature case, the matrix $A$ with the elements
\begin{equation}
  {A_{ij}} = \sqrt {\pi /2} {\left( {{i^{1/3}} + {j^{1/3}}} \right)^2} \sqrt{\frac{T_i}{i}}
\end{equation}
of rank 3 can be used as an upper bound to the kernel $C$ (\ref{eq-hatc}):
\begin{equation}\label{ac-eq}
  C_{ij} \leqslant {A_{ij}} + {A_{ji}}.
\end{equation}
Rejection sampling can be used for the cases, where there is an overestimate of collision rates (line \ref{rej-line}), which happens with the probability $1 - \frac{\sqrt{T_i/i + T_j/j}}{\sqrt{T_i/i} + \sqrt{T_j/j}} \leqslant 1 - 1/\sqrt{2} < 0.3$. The approximation is done in advance, so there is no need to recalculate it.

The update of the full structure again requires $O(R \log M)$ operations (line \ref{upds-line} of the algorithm below) with $R = 3$ for the ballistic kernel. To pick a pair of particles, a rank 1 matrix is taken with the probability, proportional to the sum of its elements ($O(R)$ operations, lines \ref{r2c-line}-\ref{r2cend-line}). Then the element of $\vec u^k$ is picked with the probability, proportional to its value ($O(\log M)$ operations through binary search in a segment tree), which determines a row $i$; then similarly an element of $\vec v^k$ is picked, which determines the column $j$ (also $O(\log M)$ operations, line \ref{ic-line}). So the total complexity is again dominated by the update procedure and is $O(R \log M)$.

The full algorithm with the corresponding modifications is presented below:

\paragraph*{Temperature-dependent Monte Carlo}
\begin{algorithmic}[1]
\REQUIRE{
Initial maximum particle size $M$.

Vector of initial numbers of particles $N_i, \quad i = \overline{1, M}$.

Vector of initial partial temperatures $T_i, \quad i = \overline{1, M}$. 

Initial volume $V = N / \sum\limits_{i = 1}^M n_i$.

Matrix $A \in \mathbb{R}^{M \times M}$ in the form $A_{ij} N_i N_j = \sum\limits_{k = 1}^R \vec u_i^k \otimes \vec v_j^k, \quad \vec u^k, \vec v^k \in {\mathbb{R}^M}$, such that $C_{ij} \leqslant \tilde C_{ij} = A_{ij} + A_{ji}$. 

Final time $maxtime$.

}
\ENSURE{$T_i$ is equal to the temperature of the size-$i$ particles at time $t = maxtime$.

$n_i = N_i/V$ is equal to the number density of the size-$i$ particles at time $t = maxtime$.
}

\STATE \COMMENT{Set initial number of particles $N_{0}$:}
\STATE $N_{0} := \sum\limits_{i = 1}^M {N_i}$
\STATE \COMMENT{Set current number of particles $N_p$ equal to initial number of partiles:}
\STATE $N := N_{0}$
\STATE Create segment trees $u_t^k$ and $v_t^k$ based on vectors $\vec u^k$ and $\vec v^k$
\STATE $curtime := 0$
\WHILE{$curtime < maxtime$}
  \REPEAT
    \STATE \COMMENT{First elements of segment trees contain the total sum, which can be used to compute the total aggregation rate:}
    \STATE $total\_rate := \sum\limits_{k = 1}^R (u_t^k)_1 (v_t^k)_1$
    \STATE \COMMENT{Update current time:}
    \STATE $curtime := curtime - 2 V \ln \left( {\rm rand} (0,1] \right) / total\_rate$
    \STATE \COMMENT{Choose a rank 1 component:} \label{r2c-line}
    \STATE $r := total\_rate \cdot {\rm rand}(0,1]$
    \FOR{$k := 1$ \TO $R$}
      \STATE \COMMENT{First elements of segment trees contain the total sum, which is subtracted from $r$:}
      \STATE $r := r - (u_t^k)_1 (v_t^k)_1$
      \IF {$r \leqslant 0$}
        \STATE \textbf{break}
      \ENDIF
    \ENDFOR \label{r2cend-line}
    \STATE Choose sizes $i$ and $j$ through binary search in segment trees $u_t^r$ and $v_t^r$ \label{ic-line}
    \STATE \COMMENT{Rejection sampling:} \label{rej-line}
    \IF{$\tilde C_{ij} \cdot {\rm rand}(0, 1] > C_{ij}$}
      \STATE \textbf{continue}
    \ENDIF
    \STATE \COMMENT{Reject collision with itself:}
    \IF{($i = j$) \AND ($N_i \cdot {\rm rand}(0,1] \leqslant 1$)} 
      \STATE \textbf{continue}
    \ENDIF
  \UNTIL{$i$ and $j$ are chosen}
  \IF{$rand(0,1] \leqslant P(i, j, T_i, T_j)$}
    \STATE \COMMENT{Restitution:}
    \STATE update $T_i$ and $T_j$ using equations (\ref{restemp-eq})
  \ELSE
    \STATE \COMMENT{Aggregation:}
    \IF{$i + j > M$}
      \STATE $M := 2M$
      \STATE Increase the sizes of all the arrays correspondingly and initialize new values by $0$
    \ENDIF
    \STATE update $T_i$, $T_j$ and $T_{i+j}$ using equations (\ref{aggtemp-eq}) \label{upds-line}
    \STATE $N_i := N_i - 1$
    \STATE $N_j := N_j - 1$
    \STATE $N_{i+j} := N_{i+j} + 1$
    \STATE $N := N - 1$
  \ENDIF
  \STATE \COMMENT{Number of particles doubling:}
  \IF{$N \leqslant N_{0}/2$}
    \STATE $N := 2N$
    \STATE $V := 2V$
  \ENDIF
  \STATE Update segment trees
\ENDWHILE
\end{algorithmic}

\subsection{Numerical examples}

\begin{figure}[ht]
\begin{subfigure}[b]{0.48\textwidth}
\centering
\includegraphics[width=\columnwidth]{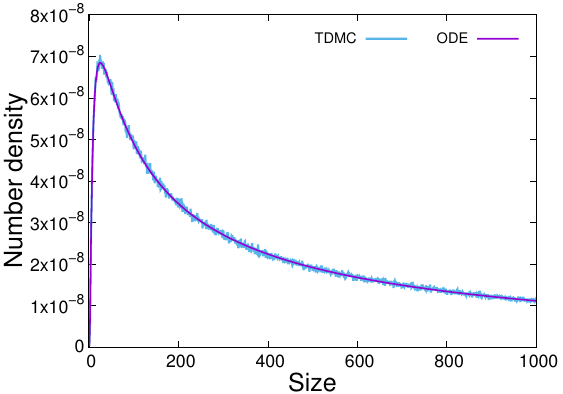}
\caption{Number density distribution.}
\end{subfigure}
\begin{subfigure}[b]{0.48\textwidth}
\centering
\includegraphics[width=\columnwidth]{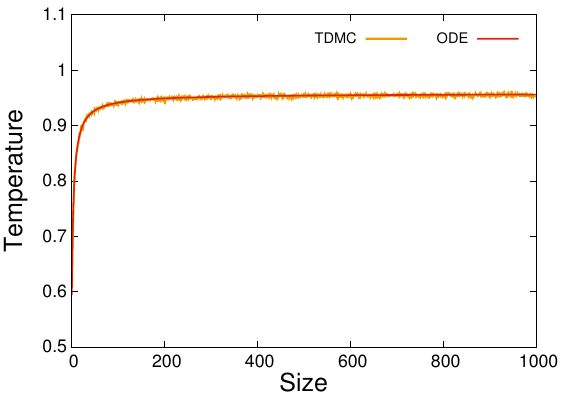}
\caption{Temperature distribution.}
\end{subfigure}
\caption{Number density and temperature for masses up to 1000 at $t=100000$. Temperature-dependent model parameters and ODE solution are taken from \cite{gen-smol}. Temperature-dependent Monte Carlo (TDMC) simulations started from $N = 10^7$ monomers.}
\label{fig-num-dens}
\end{figure}

To check the new approach, initial conditions and $C$, $B$ and $D$ kernels in \eqref{n-eq}, \eqref{nt-eq} are defined as in \cite{nature}, where the authors study temperature-dependent ballistic aggregation. The exact values of the kernels are available in \ref{mod:app}. Note that by construction the temperature-dependent Monte Carlo method should converge to the solution of temperature-dependent ODE system as the number of particles is increased (while keeping total number density the same). Figure \ref{fig-num-dens} shows number densities obtained by Monte Carlo method and numerical solution of the temperature-dependent ODE. The same number density distribution up to stochastic noise can be observed. The disappearance of small particles is a consequence of the ballistic kernel scaling, which corresponds to Case III in the conventional classification of aggregating regimes \cite{scaling}.

Though Direct Simulation Monte Carlo (DSMC), where each particle has its own speed, is closer to reality since it does not require Maxwell distribution assumption, it takes much longer to compute. The reason is that, in general, one needs to save, update and choose speeds of the colliding particles, while here speeds are already incorporated into the partial temperatures. Moreover, in practice, even in driven granular gases with exponential velocity tail, the solution of the temperature-dependent system closely matches the results of the DSMC simulations \cite{bodrova}. 

\begin{table}[ht]
\caption{Monte Carlo computation time for $10^7$ particles, temperature-dependent ballistic kernel \cite{nature}.}
\label{tab-temp}
\vspace{-0.3cm}
\begin{center}
\small
\begin{tabular}{|c|c|c|c|c|c|}
\hline
System time, 
$t$ & $M$ & Inverse, sec & Acceptance-rejection, sec & Low-rank, sec & Collisions \\
\hline
1000 & 128 & 1099 & 1188 & 1068 & $1.2 \cdot 10^9$ \\
\hline
10000 & 4096 & 95871 & 4242 & 2879 & $2.6 \cdot 10^9$ \\
\hline
100000 & 131072 & -- & 14017 & 4829 & $3.9 \cdot 10^9$ \\
\hline
1000000 & 2097152 & -- & 56213 & 7595 & $5.2 \cdot 10^9$ \\
\hline
\end{tabular}
\end{center}
\end{table}

The method is compared with the inverse and acceptance-rejection method in table \ref{tab-temp}. In the acceptance-rejection method the inequality
\begin{equation}\label{eq-balineq}
  \max\limits_{ij} \left( i^{1/3} + j^{1/3} \right)^2 \sqrt{\frac{T_i}{i} + \frac{T_j}{j}} \leqslant \left( 1 + M^{1/3} \right)^2 \sqrt{1 + \frac{1}{M}} \sqrt{\max\limits_i T_i} = C_{\max}
\end{equation}
is used to determine the bound on the maximum value of ballistic kernel elements and calculate the probabilities \eqref{eq:accrej}. 

The low-rank approach can be seen to depend logarithmically on the maximum particle size, while the inverse method has a linear dependence on the maximum size. When the maximum size reaches thousands, the low-rank approach becomes 30 times faster than the inverse method. Acceptance-rejection method is much better than the inverse, but also becomes slower than low-rank after maximum size $M$ reaches thousands.

Using low-rank Monte Carlo method allowed constructing a full phase diagram of possible behaviors in temperature-dependent Smoluchowski equation \cite{diagram}. It was also recently used to confirm the exact analytical solutions of temperature-dependent Smoluchowski equations \cite{meexact}.

\section{Low-rank DSMC}\label{sec-dsmc}

Finally, the same idea can be realised in Direct Simulation Monte Carlo too, where each particle has its own speed $\vec V = \left( V_x, V_y, V_z \right)$. Uppercase letters will be used for 3D vectors so that the notion does not conflict with $M$-dimensional vectors.

The difference between DSMC and regular Monte Carlo is that previously there was no distinction between particles of the same size. In the case of ballistic trajectories, the collision frequency between the $k$-th particle of size $i$ with speed $\vec V_i^k$ and the $l$-th particle of size $j$ with speed $\vec V_j^l$ along the random direction is
\begin{equation}\label{eq-cdsmc}
  C_{ij}^{kl} = \frac{\pi r^2}{V} \left( i^{1/3} + j^{1/3} \right)^2 \left| \left( \vec V_i^k - \vec V_j^l \right) \cdot \vec e \right|,
\end{equation}
where $r$ is the radius of the monomer, $V$ is the system volume and $\vec e$ is the collision direction \cite{dsmcbest}.

As previously, we firstly construct a simple upper bound $\tilde C$ such that
\[
  C_{ij}^{kl} N_i N_j \leqslant \tilde C_{ij} N_i N_j = A_{ij} + A_{ji}.
\]
In the case of $C$ described by the equation (\ref{eq-cdsmc}) one can use
\begin{equation}\label{eq-adsmc}
  A_{ij} = \frac{\pi r^2}{V} \left( i^{1/3} + j^{1/3} \right)^2 \max\limits_{k} \left| \vec V_i^k \right| N_i N_j.
\end{equation}
The value of $\max\limits_{k} \left| \vec V_i^k \right|$ can be increased whenever the particle exceeds the previous maximum speed. One can also update the maximum after sufficiently many collisions just by checking the speeds of all particles.

If $A$ is defined as in equation (\ref{eq-adsmc}), then
\[
  \tilde C_{ij} = \frac{\pi r^2}{V} \left( i^{1/3} + j^{1/3} \right)^2 \left( \max\limits_{k} \left| \vec V_i^k \right| + \max\limits_{l} \left| \vec V_j^l \right| \right)
\]
and the acceptance criterion
\[
  C_{ij}^{kl} > {\rm rand} (0,1] \cdot \tilde C_{ij} 
\]
becomes
\begin{equation}\label{eq-rdsmc}
  \left| \left( \vec V_i^k - \vec V_j^l \right) \cdot \vec e \right| > {\rm rand} (0, 1] \cdot \left( \max\limits_{k} \left| \vec V_i^k \right| + \max\limits_{l} \left| \vec V_j^l \right| \right).
\end{equation}
Vector $\vec e$, as well as particles $k$ and $l$ can be selected at random.

Therefore, the choice of colliding particles is performed in two steps. Firstly, the sizes are selected according to the matrix $A$ (\ref{eq-adsmc}), which is done similarly to previous sections. Next, the direction $\vec e$ and speeds $\vec V_i^k$ and $\vec V_j^l$ are selected at random. If they are rejected according to equation (\ref{eq-rdsmc}), one should go back to choosing the sizes again.

After choosing the particles, it should be determined whether they aggregate with some probability $P(i, j, \vec V_i^k, \vec V_j^l)$ (obtained from the collision model), which now depends on their speeds.

Overall, the low-rank DSMC is performed as follows:

\paragraph*{Low-rank DSMC algorithm}
\begin{algorithmic}[1]
\REQUIRE{
Initial maximum particle size $M$.

Vector of initial numbers of particles $N_i, \quad i = \overline{1, M}$.

Initial speeds $\vec V_i^k, \quad i = \overline{1, M}, \quad k = \overline{1, N_i}$. 

Initial volume $V = N / \sum\limits_{i = 1}^M n_i$.

Matrix $A \in \mathbb{R}^{M \times M}$ in the form $A = \sum\limits_{k = 1}^R \vec u^k \otimes \vec v^k, \quad \vec u^k, \vec v^k \in {\mathbb{R}^M}$, such that $C_{ij} N_i N_j \leqslant \tilde C_{ij} N_i N_j = A_{ij} + A_{ji}$. 

Final time $maxtime$.

}
\ENSURE{Speeds $\vec V_i^k$ of the size-$i$ particles at time $t = maxtime$ for any $i$.

$n_i = N_i/V$ is equal to the number density of the size-$i$ particles at time $t = maxtime$.
}

\STATE Create segment trees $u_t^k$ and $v_t^k$ based on vectors $\vec u^k$ and $\vec v^k$
\STATE $curtime := 0$
\WHILE{$curtime < maxtime$}
  \REPEAT
    \STATE \COMMENT{First elements of segment trees contain the total sum, which can be used to compute the total aggregation rate:}
    \STATE $total\_rate := \sum\limits_{k = 1}^R (u_t^k)_1 (v_t^k)_1$
    \STATE \COMMENT{Update current time:}
    \STATE $curtime := curtime - 2 V \ln \left( {\rm rand}(0,1] \right) / total\_rate$
    \STATE \COMMENT{Choose a rank 1 component:}
    \STATE $r := total\_rate \cdot {\rm rand}(0,1]$
    \FOR{$k := 1$ \TO $R$}
      \STATE \COMMENT{First elements of segment trees contain the total sum, which is subtracted from $r$:}
      \STATE $r := r - (u_t^k)_1 (v_t^k)_1$
      \IF {$r \leqslant 0$}
        \STATE \textbf{break}
      \ENDIF
    \ENDFOR
    \STATE Choose sizes $i$ and $j$ through binary search in segment trees $u_t^r$ and $v_t^r$
    \STATE Randomly choose particles $k$ and $l$ of size $i$ and $j$ and a collision direction $\vec e$.
    \STATE \COMMENT{Rejection sampling:}
    \IF{$\tilde C_{ij}^{kl} \cdot {\rm rand}(0, 1] > C_{ij}$}
      \STATE \textbf{continue}
    \ENDIF
    \STATE \COMMENT{Reject collision with itself}
    \IF{($i = j$) \AND ($k = l$)} 
      \STATE \textbf{continue}
    \ENDIF
  \UNTIL{$i$, $j$, $k$ and $l$ are chosen}
  \IF{${\rm rand}(0,1] \leqslant P(i, j, \vec V_i^k, \vec V_j^l)$}
    \STATE \COMMENT{Restitution:}
    \STATE \COMMENT{Update $\vec V_i^k$ and $\vec V_j^l$ based on the restitution coefficient $\varepsilon \left( i, j, \vec V_i^k - \vec V_j^l, \vec e \right)$:}
    \STATE $\vec V_i^k := \vec V_i^k - \frac{1 + \varepsilon}{2} \cdot \frac{j}{i+j} \left( \left( \vec V_i^k - \vec V_j^l \right) \cdot \vec e \right) \vec e$
    \STATE $\vec V_j^l := \vec V_j^l + \frac{1 + \varepsilon}{2} \cdot \frac{i}{i+j} \left( \left( \vec V_i^k - \vec V_j^l \right) \cdot \vec e \right) \vec e$
  \ELSE
    \STATE \COMMENT{Aggregation:}
    \IF{$i + j > M$}
      \STATE $M := 2M$
      \STATE Increase the sizes of all the arrays correspondingly and initialize new values by $0$
    \ENDIF
    \STATE Remove particles $k$ and $l$
    \STATE Add a new particle of size $i + j$ and speed $\frac{i \vec V_i^k + j \vec V_j^l}{i + j}$
  \ENDIF
  \STATE Update segment trees
\ENDWHILE
\end{algorithmic}

Here the particle doubling is not performed since one cannot replace the lost particles without affecting the distribution of speeds in some way, so there is usually a direct correspondence between real and numerical particles. The real system is then modelled as a rescaled version of the small numerical system.


As an example, let us consider a system without aggregation.

For simplicity, consider collisions with the constant restitution coefficient $\varepsilon = 0.99$ and the exponential number density distribution:
\[
  N_k = \left\lfloor N_1 \exp \left( -a \cdot \left( k - 1 \right) \right) \right\rfloor.
\]
All temperatures $T_k$ are initially set to 1, and the initial speed distribution is gaussian. Number density is set to 0.1 particles per unit volume. Monomers have unit mass and diameter.

The acceptance condition in the acceptance-rejection method is \cite{dsmcbest}:
\[
  \left( i^{1/3} + j^{1/3} \right)^2 \left| \left( \vec V_i^k - \vec V_j^l \right) \cdot \vec e \right| > {\rm rand}(0, 1] \cdot \max\limits_{i,j} \left( i^{1/3} + j^{1/3} \right)^2 \left( \max\limits_k \left| \vec V_i^k \right| + \max\limits_l \left| \vec V_j^l \right| \right),
\]
where the size rejection is also accounted for. One can already see the main problem of the acceptance-rejection method: when the rejection of sizes is combined with the rejection of speeds, the total rejection probability significantly increases, which is known to be the main problem of polydisperse systems modelling \cite{dsmc}. Moreover, the exact maximum in the r.h.s. is unknown and takes much time to calculate. To avoid this problem, the comparison is made with the acceptance-rejection strategy, where collision of size $i$ and $j$ particles is accepted if
\begin{equation}\label{eq-dsmccon}
   \left( i^{1/3} + j^{1/3} \right)^2 \left| \left( \vec V_i^k - \vec V_j^l \right) \cdot \vec e \right| > {\rm rand}(0, 1] \cdot 2 \max\limits_{i} \left( i^{1/3} + M^{1/3} \right)^2 \max\limits_k \left| \vec V_i^k \right|.
\end{equation}
Since in the current system particles with larger size usually have a lower maximum speed, the lower bound on the maximum is only two times smaller than the upper bound:
\[
\begin{aligned}
\max\limits_{i,j} \left( i^{1/3} + j^{1/3} \right)^2 \left( \max\limits_k \left| \vec V_i^k \right| + \max\limits_l \left| \vec V_j^l \right| \right) & \leqslant 2 \max\limits_{i} \left( i^{1/3} + M^{1/3} \right)^2 \max\limits_k \left| \vec V_i^k \right|, \\
  \max\limits_{i,j} \left( i^{1/3} + j^{1/3} \right)^2 \left( \max\limits_k \left| \vec V_i^k \right| + \max\limits_l \left| \vec V_j^l \right| \right) & \geqslant \max\limits_{i} \left( i^{1/3} + M^{1/3} \right)^2 \max\limits_k \left| \vec V_i^k \right|,
\end{aligned}
\]
so our condition (\ref{eq-dsmccon}) differs by at most 2 times from the one, where the maximum is explicitly calculated.

The simulation times for both methods are presented in tables \ref{tab-dsmc01} and \ref{tab-dsmc001}. The low-rank approach is more efficient, and the difference is more significant, when the number of different sizes is larger. It was previously successfully utilised to study non-aggregating granular gases \cite{bodrova,bodrovamol}, where the system included $3 \cdot 10^8$ particles, and more than $10^{10}$ collisions were performed.

\begin{table}[ht]
\caption{DSMC computation time for $10^5$ particles and exponential distribution $N_k \sim \exp \left( - 0.1 \cdot k \right)$.}
\label{tab-dsmc01}
\vspace{-0.3cm}
\begin{center}
\small
\begin{tabular}{|c|c|c|c|}
\hline
System time, 
$t$ & Acceptance-rejection, sec & Low-rank, sec & Collisions \\
\hline
1 & 0.46 & 0.035 & $6.0 \cdot 10^4$ \\
\hline
10 & 4.03 & 0.34 & $5.8 \cdot 10^5$ \\
\hline
100 & 34.3 & 3.0 & $5.0 \cdot 10^6$ \\
\hline
\end{tabular}
\end{center}
\end{table}

\begin{table}[ht]
\caption{DSMC computation time for $10^5$ particles and exponential distribution $N_k \sim \exp \left( - 0.01 \cdot k \right)$.}
\label{tab-dsmc001}
\vspace{-0.3cm}
\begin{center}
\small
\begin{tabular}{|c|c|c|c|}
\hline
System time, 
$t$ & Acceptance-rejection, sec & Low-rank, sec & Collisions \\
\hline
1 & 1.17 & 0.09 & $9.2 \cdot 10^4$ \\
\hline
10 & 10.6 & 0.78 & $8.9 \cdot 10^5$ \\
\hline
100 & 79 & 5.3 & $7.1 \cdot 10^6$ \\
\hline
\end{tabular}
\end{center}
\end{table}

\section{Conclusion}\label{sec-con}

The new low-rank technique is presented, which enhances the computational performance of Monte Carlo methods in simulation of the evolution of size-polydisperse systems. Its effectiveness is illustrated for Monte Carlo simulations of traditional Smoluchowski equations and temperature-dependent Smoluchowski equations. The main advantage of the new approach is the logarithmic dependence of the computational complexity on the maximum particle size $M$, while the conventional inverse and acceptance-rejection methods show polynomial dependence. This allows our method to perform collisions more than ten times faster in case of ballistic kernel, and its advantages become more apparent with increasing system size.

\section*{Acknowledgements}
The study was supported by a grant from the Russian Science Foundation No.~21-11-00363, https://rscf.ru/project/21-11-00363/.

\appendix

\section{Temperature-dependent ballistic kernels}\label{mod:app}

Here the temperature-dependent ballistic kernels from \cite{nature, gen-smol} are specified, which were used to test the temperature-dependent Monte Carlo method.
\[
\begin{gathered}
  {C_{ij}} = 2\sqrt {2\pi } \sigma _{ij}^2\sqrt {{T_i/i} + {T_j/j}} \hfill \\
  {C_{ij}^{\rm agg}} = C_{ij} P_{ij} \hfill \\
  {P_{ij}} = 1 - f_{ij}, \hfill \\
  {B_{ij}} = 2\sqrt {2\pi } \sigma _{ij}^2\frac{1}{{\sqrt {{T_i/i} + {T_j/j}} }}\left( {\frac{T_i T_j}{ij} \left( {1 - {f_{ij}}} \right) + \frac{4}{3}{{\left( {\frac{T_i - T_j}{{i + j}}} \right)}^2}\left( {1 - {g_{ij}}} \right)} \right), \hfill \\
  {D_{ij}^{\rm agg}} = 2\sqrt {2\pi } \sigma _{ij}^2\frac{1}{\sqrt{T_i/i + T_j/j }}\left( {\frac{T_i T_j}{ij} \left( {1 - {f_{ij}}} \right) + \frac{4}{3} \left( \frac{T_i}{i} \right)^2 \left( {1 - {g_{ij}}} \right)} \right), \hfill \\
  D_{ij}^{\rm res} = \frac{4j}{3 \left( i + j \right)}\sqrt {2\pi } \sigma _{ij}^2\sqrt {T_i/i + T_j/j} \left( \frac{T_i}{i} \left( 1 - \varepsilon^2 \right) g_{ij} +  \frac{T_i - T_j}{i+j} \left( 1 + \varepsilon \right)^2 g_{ij} \right), \hfill \\
  {f_{ij}} = {e^{ - {q_{ij}}}}\left( {1 + {q_{ij}}} \right), \hfill \\
  {g_{ij}} = {e^{ - {q_{ij}}}}\left( {1 + {q_{ij}} + q_{ij}^2/2} \right), \hfill \\
  {q_{ij}} = \frac{{{W_{ij}}}}{{{\varepsilon ^2}\tfrac{{ij}}{{i + j}}\left( T_i/i + T_j/j \right)}}, \hfill \\
  {W_{ij}} = a \left( \frac{i^{1/3} j^{1/3}}{\left( i^{1/3} + j^{1/3} \right)} \right)^{\Lambda}, \hfill \\
  {\sigma _{ij}} = \frac{{{i^{1/3}} + {j^{1/3}}}}{2}. \hfill \\
\end{gathered}
\]
Here $\sigma_{ij}$ is the collision cross-section for unit diameter of monomers. Aggregation barrier $W_{ij}$ is the amount of kinetic energy needed to avoid aggregation. $a$ describes the strength of the barrier, while $\Lambda$ is the homogeneity coefficient. $\varepsilon$ is the restitution coefficient. The parameters and initial conditions are the same as in \cite{gen-smol}:
\[
\begin{gathered}
  \varepsilon  = 0.99, \hfill \\
  a = 0.1, \hfill \\
  \Lambda = 0.4, \hfill \\
\end{gathered}
\]
initial conditions are monodisperse $n_1 \left(t = 0 \right) = 0.3/\pi$ with $T_1 = 1$.

\bibliography{mybibfile}


\end{document}